\documentclass[agupp]{aguplus}
\usepackage[reqno]{amsmath}
\usepackage{times}
\usepackage{graphicx}
\usepackage{float}
\usepackage{psfrag}
\usepackage{verbatim}
\lefthead{NEWBURY ET AL.}
\righthead{SHOCK WIDTH MEASUREMENT}
\authoraddr{M. Gedalin, Department of Physics,
Ben-Gurion University, P.O. Box 653, Beer-Sheva 84105, Israel. 
(e-mail: gedalin@bgumail.bgu.ac.il); J. A. Newbury and C. T. Russell, 
IGPP/UCLA, 405 Hilgard Ave., Los Angeles, CA 90095-1567 (e-mail: 
newbury@igpp.ucla.edu, ctrussell@igpp.ucla.edu)}
\begin{document}
\title{The determination of shock ramp width using the noncoplanar 
magnetic field 
component}
\author{J. A.  Newbury and C. T.  Russell}
\affil{Institute of Geophysics and 
Planetary Physics and the Department of Earth and Space Sciences, 
University of
California Los Angeles} 
\author{M.  Gedalin} 
\affil{Department of 
Physics, Ben-Gurion University, Beer-Sheva, Israel}
\begin{abstract}
We determine a simple expression for the ramp width of a 
collisionless fast
shock, based upon the relationship between the noncoplanar  
and main magnetic field components. By comparing this predicted width 
with
that measured during an observation of a shock, the shock velocity can
be determined from a single spacecraft. For a range of low-Mach, 
low-$\beta$ bow shock observations
made by the ISEE-1 and -2 spacecraft, ramp widths determined from 
two-spacecraft
comparison and from this noncoplanar component relationship agree 
within 30\%.
When two-spacecraft measurements are not available or are 
inefficient, this technique 
provides a reasonable estimation of scale size for low-Mach shocks.
\end{abstract}

\section{Introduction}

The determination of spatial scales within the collisionless shock 
front is one of the central problems 
for observational shock physics. The width of the shock ramp, 
defined as the main 
transition layer between upstream to downstream plasma, is of
particular interest. Without spacecraft measurements in a spatial 
rather 
then temporal frame of reference, it is impossible to make any 
comparisons 
between observations and theoretical models. 

For bow shock studies, generally one of two methods has been applied 
to 
transform the time 
series observed by in-situ magnetometers
into a spatial magnetic field profile: (1) the comparison of shock
observations made by multiple spacecraft with known separations 
in time and space 
\cite[for example,][]{Ru82,Fa93,NR96}, and (2) the comparison of the 
temporal duration of the shock foot feature as observed by a single 
spacecraft 
with the spatial foot length predicted by a model based
on the motion of specularly reflected ions 
\cite[for example,][]{Sc83,GT85,NR96}. Both methods assume that 
the bow shock is stationary and 
one-dimensional, and each has its own limitations. The first is not 
reliable when the 
time delay between spacecraft observations is too large \cite[]{NR96} 
(non-stationarity can affect the results) or too small 
(relative errors become large). Also, large transverse 
spatial separations of the spacecraft
with respect to the shock front can introduce error due 
to the three-dimensional nature of the
bow shock. The second method is not applicable to laminar shocks 
(shocks
observed during low Mach number and low plasma 
$\beta$ conditions). Ion reflection does not
play a dominant dissipative role at such shocks, and so there is 
little or no
foot structure available to measure. (In high Mach number shocks, the second 
method
can also be problematic, since the ion reflection is clearly not 
specular 
\cite[]{GT85,Ge96c} which affects the model's predictions.) 

Because of these limitations, it is desirable to have another 
independent 
method for determining shock scale lengths, particularly for laminar 
shock observations
made by a single spacecraft. In this paper, we make use of the 
noncoplanar component of
the magnetic field in the shock ramp in order to estimate a scale 
size 
which can in turn be compared with spacecraft observations of low 
Mach number shocks. 
In Section 2 we briefly outline the theoretical basis of this method, 
and then apply 
the technique to an example of a bow shock observation made by the 
ISEE spacecraft (Section 3).
In Section 4, we examine the results obtained from a variety of low 
Mach and low $\beta$
bow shock observations, and find that the proposed method works well, 
even for shocks which
are not strictly laminar.

\section{Theoretical Basis}

Within the ramp layer of a fast collisionless shock (such as the 
Earth's bow
shock), the magnetic field is observed to rotate out of the 
coplanarity plane
(the plane defined by the shock normal and the upstream and 
downstream magnetic field
vectors) \cite[]{Th87}. 
The analytical relation between this noncoplanar component and the 
main 
magnetic component of the shock profile was first derived by 
\cite{JE87} phenomenologically in an
integral form; its approximate nature has been shown observationally 
by \cite{Go88,Fr90}.  Recently, \cite{Ge96a} examined the noncoplanar 
relation 
using a general two-fluid hydrodynamics approach, and carried out the 
derivation
with only the widely accepted assumptions of shock stationarity, 
one-dimensionality, 
and quasi-neutrality. In the coplanarity coordinate system where $N$ 
denotes the direction along the shock normal,
$L$ is along the magnetic field component in the shock plane, and $M$ is directed 
out of the coplanarity
plane, the general expression for the noncoplanar magnetic field 
component ($B_M$) is as follows:
\begin{equation}\label{general}
\begin{split}
& B_M(1-\frac{B_N^2}{4\pi n m_iv^2})=\frac{cB_N}{4\pi 
nve}\frac{d}{dN}B_L\\
& 
-\frac{c}{nve}\frac{d}{dN}P_{NL}^{(e)}+\frac{B_N}{nm_iv^2}(P_{NM}^{(e)} 
+P_{NM}^{(i)})
\end{split}
\end{equation}
in the limit $m_e\rightarrow 0$, and where $n$ is the number density; 
$v$ is the $N$ component 
 of the hydrodynamic velocity; $B_N=\text{const}$ and 
$nv=\text{const}$ (number flux conservation); 
and $P_{ij}$ are components of the 
pressure tensor.  

 It has been shown by \cite{GZ95} that the appearance of $P_{NM}$ is 
 mainly
due to the presence of reflected ions, and that $P_{NM}\ll 
n_um_iV_u^2$ 
for low Mach number shocks \cite[]{Ge96b}. For laminar shocks it 
is expected \cite[]{Th85} that the relative contribution of the 
pressure terms is 
$\sim \beta/M^2$ l and therefore the following approximate 
 expression holds:
 \begin{equation}
 B_M=l_W\frac{dB_L}{dN},\label{final}
 \end{equation}
 where $l_W=c \cos \theta_{BN}/M\omega_{pi}$ (i.e., $k = 
1/l_W$ would be the wavenumber of 
a whistler, phasestanding  upstream of the ramp), $\theta_{BN}$ is the angle 
between the
 upstream magnetic field and the shock normal, $M_A$ is the 
Alfv\'{e}nic Mach number, and 
 $c/\omega_{pi}$ is the ion intertial length. It should be emphasized 
that 
 \eqref{final} is a differential analog of 
 the integral relation developed by \cite{JE87}. However, 
\eqref{final} should be applied 
 only within the ramp where the main contribution $\propto dB_L/dN$ 
is 
 not small, otherwise the term $\propto P_{NM}$ can dominate.

Therefore, by estimating the slope of the main magnetic field 
component
($dB_L/dt$) in the vicinity of the maximum noncoplanar
component in an observed shock ramp, one can determine
an independent estimate of the velocity of the shock front from 
\eqref{final}. 
It is then straightforward to transform the observed 
temporal shock profile into a spatial one suitable for comparison 
with theory, other
shock observations, and/or simulations. 
An advantage of the method outlined above is that it gives the scale 
in the 
relative physical units and not absolute units. It encounters, 
however, the 
same  difficulties from the determination of the vector basis 
$(N, M, L)$, as all other methods.  

Further difficulties can arise from \eqref{final} since it is 
sensitive to gradients in the field profile; noise and wave activity 
associated with an
observation of a bow shock can make the measurement of $dB_L/dt$ 
difficult, and traditional filtering techniques do not preserve 
gradients well.
For laminar shocks, a simple variation of the method approximates
the shock ramp profile with a hyperbolic tangent function along the 
$N$ direction:
\begin{equation}
B_L=\frac{B_u+B_d}{2}+\frac{B_d-B_u}{2}\tanh\frac{3N}{l_r}, 
\label{model}
\end{equation}
where $B_u$ and $B_d$ refer to the total field upstream and 
downstream of
the shock front, and the coefficient 3 ensures that $90\%$ of the 
magnetic field 
variation occurs within the ramp, $-l_r/2<N<l_r/2$. In this case, 
from \eqref{final} one 
immediately has an expression for the ramp width:
\begin{equation}
l_r=1.5l_W\frac{B_d-B_u}{B_{M,\text{max}}}. \label{model2}
\end{equation}  
This approach requires accurate measurements of $B_u$ and $B_d$, but 
is less sensitive 
to the local $B_L$ gradient than the direct application of 
\eqref{final} to an observed
shock profile. 

\section{A Sample Bow Shock}

\begin{figure}[htb]
\psfrag{time}[][]{$$}
\psfrag{bt}[b][t]{$B_t$}
\psfrag{bn}[b][t]{$B_N$}
\psfrag{bm}[b][t]{$B_M$}
\psfrag{bl}[b][t]{$B_L$}
\includegraphics[scale=0.45]{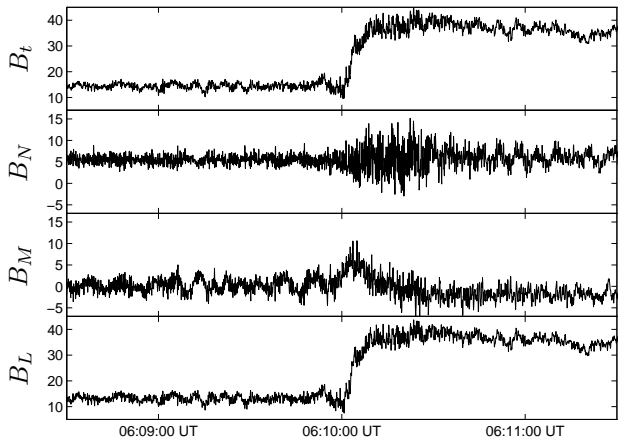}
\caption{The magnetic profile in the coplanar frame for a laminar shock
observed by ISEE-1 on November 26, 1977. $M_A=2.7$, $\beta=0.52$,
and $\theta_{Bn}=67.0^\circ$.}
\label{s0066nif}
\end{figure}

In the present section we apply the proposed method to a 
quasi-perpendicular collisionless shock crossing that was observed by 
the
ISEE-1 and -2 spacecraft on November 26, 1977, 06:10 UT. 
The magnetic profile of this shock was measured by the ISEE-1 and -2 
UCLA fluxgate
magnetometers \cite[]{Ru78}. Data is filtered to obey the Nyquist 
criterion and then sampled
at the rate of 16 vectors per second. By averaging over a minute 
of data upstream
and downstream of the shock front and applying the coplanarity 
theorem, the shock normal
is determined, and the angle between the shock 
normal and the upstream magnetic field is found to be 
$\theta_{BN}=67^\circ$. 
Figure~\ref{s0066nif}
shows the high-resolution ISEE-1 observation of total magnetic field 
and its 
three components $(B_N,B_M,B_L)$, rotated into coplanar coordinates. 

Plasma measurements of the upstream solar wind are obtained from the 
ISEE-1 and ISEE-3 solar wind
experiments \cite[]{Ba78}, and are used to calculate the following 
parameters:
 ion inertial length, $c/\omega_{pi}=58\,\text{km}$;
the Alfv\'{e}nic Mach number $M_A=2.7$  (so that $l_W= 
8.3\,\text{km}$); 
and electron and ion beta, $\beta_e=0.36$ and $\beta_i=0.16$, 
respectively. 

\begin{figure}[htb]
\psfrag{time}[][]{}
\psfrag{bt}[b][t]{$B_t$}
\psfrag{bn}[b][t]{$B_N$}
\psfrag{bm}[b][t]{$B_M$}
\psfrag{bl}[b][t]{$B_L$}
\includegraphics[scale=0.45]{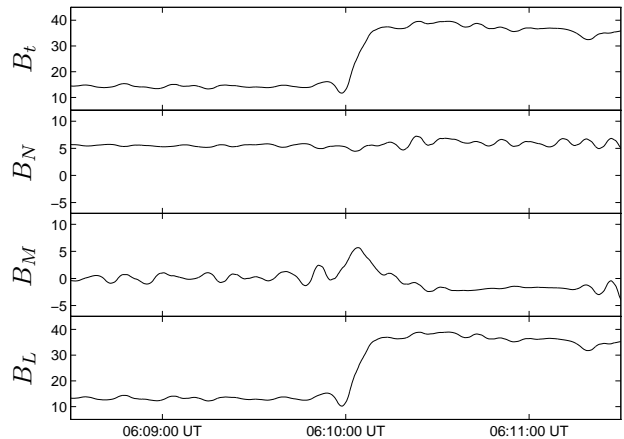}
\caption{The magnetic profile of the shock, with short period noise
removed using a wavelet filter (Daubechies, 10 wavelet with the 6 
finest scales removed).}
\label{s0066nifdn6}
\end{figure}

In order to remove short wavelength noise while maintaining the 
gradients within
the shock profile, the data was denoised applying a 
discrete wavelet transform   
\cite[see, for example,][]{Ch92,Do93}
(using the Daubechies-10 wavelet) and removing the 6 finest scales. 
This approximately
corresponds to the removal of scales shorter 
than $2^6=64$ data points, which in turn roughly corresponds to  
4 second averaging for this 1/16 sec. resolution observation.
Although substantial oscillations persist in the upstream
and downstream regions, the behavior of 
$B_M$ and $B_L$ within the ramp is consistent with the theoretical 
prediction, 
as seen in Figure~\ref{s0066nifdn6}.

 Comparison of 
$B_{M,\text{max}}$ with the slope of $B_L$ gives the following 
 shock velocity estimate from the noncoplanarity:
$V_{\text{sh}}=4.4\,\text{km}/\text{s}$.      
Independently, the shock velocity estimated from the 
ISEE-1 and ISEE-2 spacecraft separation is 
$V_{sh}=5.7\,\text{km}/{s}$ (separation $L_s=20\,\text{km}$ 
along the shock normal and ramp crossing time separation of 3.5 s). 
The two 
estimates agree within 50\%  deviation. 

Applying the hyperbolic tangent approximation from equation 
\eqref{model}, the ramp
width is estimated to be $47.72 \pm 7.5\,\text{km}$. 
Measuring the ramp based upon two-spacecraft
comparisons, ramp width is found to be $56.7 \pm 8.2 \,\text{km}$.  (The 
temporal duration of the ISEE-1
ramp observation is approximately 10.25 sec.) Uncertainty in both 
calculations
is primarily dominated by the
uncertainty in the shock normal determination (which is in turn a 
result in deviations
from the upstream and downstream field measurements). These two 
estimations of ramp
width agree within 20\%.

\section{Application to a Variety of Shocks}

In order to estimate the reliability of the method outlined in the 
previous
section, we compare the results of the proposed approach
from a variety of low-Mach number shock observations. Table 1
contains relevant parameters for 
a selection of shocks observed by the 
ISEE spacecraft: the Alfvenic Mach number ($M_A$), ratio of 
criticality ($R_c$), $\theta_{BN}$ (as determined by coplanarity), 
total $\beta$ of the 
upstream plasma, the variation of the normal component of the 
magnetic 
field during the observation of the shock ramp (normalized by the 
maximum 
noncoplanar component in the ramp, $\Delta B_N/B_{M,\text{max}}$), 
and 
the results from estimating ramp width using equation \eqref{model} 
($l_{r,\text{pred}}$) 
and from the comparison of the ISEE observations ($l_{r, 
\text{obs}}$). 
These shocks were selected for their low-$\beta$, low-Mach number, 
quasi-perpendicular ($\theta_{BN} > 45^\circ$) characteristics, as 
well as being observed 
at times when the ISEE-1 and -2 spacecraft configurations were ideal 
for determining 
fine spatial scales (i.e., small spatial and temporal separations 
between observations, and 
$\theta_{BN}$'s calculated by coplanarity and by using an ellipsoidal 
model of 
the bow shock agree within $10^\circ$). Nearly-perpendicular shocks 
($\theta_{BN} > 80^\circ$) are avoided due to the difficulties of
determining the shock normal vectors from coplanarity for such 
shocks. Also, perpendicular 
shocks may not have the same whistler mode structure as shocks with 
lower 
$\theta_{BN}$ \cite[]{NR96,Fr90}. Many of these shocks have been 
examined previously by \cite{Fa93}.

For six out of seven shocks, agreement between observations and predictions is very 
good (within 30\%); in comparison, a study of shock velocities
based upon two spacecraft observations and on estimations of shock foot 
length reports agreement within 35\%   
for only half of the shocks examined \cite[]{LL94}.
Also, observation and prediction using the method outlined in this paper are 
comparable even when the shock is no longer strictly laminar: 
several 
of the shocks listed in Table~1 are slightly supercritical (i.e., the 
ratio of criticality is greater than 1) and are associated with a 
plasma 
$\beta$ that isn't especially low ($\beta > 0.3$). 

\begin{figure}[htb]
\psfrag{var}[t][b]{\small $\Delta B_N/B_{M,\text{max}}$}
\psfrag{log}[b][t]{\small $\log(l_{r,\text{obs}}/l_{r,\text{pred}})$}
\includegraphics[scale=0.65]{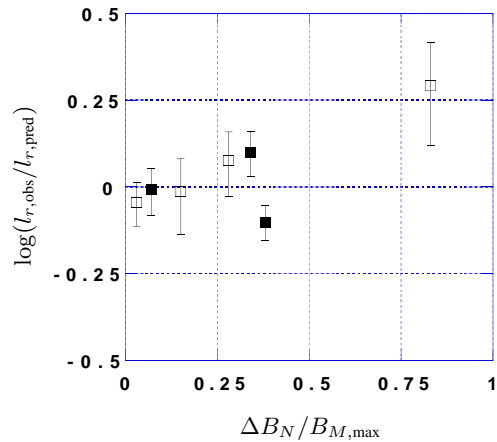}
\caption{The agreement between observation and prediction of ramp width
vs. the deviation of the normal magnetic field component within the ramp 
observation. When $B_N$ remains constant throughout the shock ramp, the 
prediction of ramp width is the most accurate.  Open squares indicate 
supercritical shocks.}
\label{stat} 
\end{figure}

The shock observed on 79 August 13, 1427 UT is an exception:  
equation \eqref{model} 
does not accurately estimate its ramp width. This can be explained by 
considering the effects of turbulence as evidenced in the deviations 
of the $B_N$ field 
component within the ramp layer. Equation \eqref{final} assumes $B_N$ 
to remain constant 
throughout the shock observation, but in reality this is not always 
so. 
Two dimensional disturbances and plasma turbulence on the shock front 
can obscure 
the coplanarity rotation. Within the shock ramp on 79 Aug. 13, 
fluctuations of $B_N$ are on 
the order of $B_{M{,max}}$, resulting in an 
under-estimated ramp width from \eqref{model}. Even with the somewhat 
stringent requirements
placed on selecting the shocks listed in Table 1, non-stationarity 
and turbulence are
still a factor which cannot be ignored. In Figure~\ref{stat},
the ratio of observed and predicted ramp widths are compared with the 
deviation of 
$B_N$ (normalized to the maximum $B_M$ component).  The shock 
ramps that agree best with the prediction also have the most constant 
$B_N$ 
components, and even a noticeable deviation in $B_N$ can still result 
in a 
reasonable estimation of ramp width. The light data points in 
Figure~\ref{stat}
correspond to supercritical shocks ($R_c > 1$).

\section{Conclusions}
We have examined the relationship between the noncoplanar component 
and gradient of the
main component of the magnetic field within the collisionless shock 
ramp layer.
By estimating the scale size of the ramp width based upon this 
relationship and comparing that 
length with the temporal duration of a shock ramp observation, the 
velocity of the shock speed
in the spacecraft frame can be calculated.
The observed temporal shock profile can then be transformed into a 
spatial frame, suitable for comparison with other shock observations 
and with theory. Based upon a
sampling of bow shock observations made by the ISEE-1 and -2 
spacecraft, we conclude that 
this technique is a 
satisfactory alternative when two-spacecraft comparisons are not 
feasible, provided that Mach number and plasma $\beta$ are low 
(although not strictly 
laminar) and the rotation of the shock profile into the coplanarity 
plane is fairly clean.

  \acknowledgements
 The wavelet transform  has been done using WaveLab package for 
Matlab. 
Magnetic field plots has been produced using Matlab. 
This research was supported by grant  94-00047 from the United 
States-Israel Binational Science Foundation (BSF), Jerusalem, Israel.

\newpage
\begin{table}[tbp]
	\centering
	\caption{Shock Parameters, Normal Component Variation in the Ramp, 
and 
Ramp Width Calculations}
\begin{tabular}{|c|c|c|c|c|c|c|c|c|}
\hline
Date &	$M_A$&	$R_c$&	$\theta_{BN}$&	$\beta$&	
$\frac{\Delta{B_N}}{B_{M,\text{max}}}$&	
$l_{r, \text{pred}}$ [km]&	$l_{r, \text{obs}}$ [km]&	
$l_{r, \text{obs}}/l_{r, \text{pred}}$\\
\hline
\small 77 Nov 26, 0610 UT&	2.73&	1.16&	$67.0^\circ$&	0.52&	0.28&	
47.7 $\pm$ 7.5 
&	56.7 $\pm$ 8.2 	
&1.19 $\pm$ 0.25\\
\small 77 Nov 26, 0619 UT&	3.07&	1.32&	$69.3^\circ$&	0.62&	0.15&
	52.3 $\pm$ 9.1 &
	50.6 $\pm$ 8.6	& 0.97 $\pm$ 0.24\\
\small 78 Aug 27, 2007 UT&	2.23&	0.85&	$74.6^\circ$&	0.16&	0.38&	
127 $\pm$12 &
	100.3 $\pm$ 5.6&	0.790 $\pm$ 0.09\\
\small 79 Aug 13, 1427 UT&	3.75&	1.4&	$78.7^\circ$&	0.10&	0.83&
	34 $\pm$ 11&
	66.7 $\pm$ 2.4	&1.96 $\pm$ 0.64\\
\small 79 Sep 18, 1029 UT&	2.92&	1.15&	$62.3^\circ$&	0.18&	0.03&
	92 $\pm$ 13&
	83.2 $\pm$ 1.5	&0.90 $\pm$ 0.13\\
\small 80 Sep 6, 1006 UT&	2.44&	0.98&	$61.6^\circ$&	0.17&	0.34&
	71.5 $\pm$ 
9.3	&
90.0 $\pm$ 6.6	&1.26 $\pm$0.19\\
\small 80 Dec 19, 1435 UT	&1.67&	0.62&	$74.8^\circ$&	0.04&
	0.07&	102 $\pm$ 
15&
	100.0 $\pm$ 5.8&	0.98 $\pm$ 0.15\\
	\hline
	\end{tabular}
	\label{table}
\end{table}

\end{document}